\documentclass[a4paper,11pt]{article}
\usepackage{jinstpub}
\usepackage{multirow}
\usepackage{bm}

\title{Systematic errors in tokamak magnetic equilibrium reconstruction: a study of EFIT++ at tokamak COMPASS}

\author[a,b,1]{K. Jirakova,\note{Corresponding author.}}
\author[a,c]{O. Kovanda,}
\author[a]{J. Adamek,}
\author[a]{M. Komm}
\author[a]{and J. Seidl}

\affiliation[a]{Institute of Plasma Physics, Czech Academy of Sciences, Za Slovankou 3, 182 00 Prague, Czech Republic}
\affiliation[b]{Faculty of Nuclear Sciences and Physical Engineering, Czech Technical University, Brehova 7, 115 19 Prague, Czech Republic}
\affiliation[c]{Faculty of Mathematics and Physics, Charles University, Ke Karlovu 3, 121 16 Prague, Czech Republic}

\emailAdd{jirakova@ipp.cas.cz}

\abstract{Uncertainties and errors in magnetic equilibrium reconstructions are a wide-spread problem in interpreting experimental data measured in the tokamak edge. This study demonstrates errors in EFIT++ reconstructions performed on the COMPASS tokamak by comparing the outer midplane separatrix position to the Velocity Shear Layer (VSL) position. The VSL is detected as the plasma potential peak measured by a reciprocating ball-pen probe. A subsequent statistical analysis of nearly 400 discharges shows a strong systematic trend in the reconstructed separatrix position relative to the VSL, where the primary factors are plasma triangularity and the magnetic axis radial position. This dependency is significantly reduced after the measuring coils positions as recorded in EFIT input are optimised to provide a closer match between the "synthetic" coil signal calculated by the Biot-Savart law in a vacuum discharge and the actual coil signal. In conclusion, we suggest that applying this optimisation may lead to more accurate and reliable reconstructions of the COMPASS equilibrium, which would have a positive impact on the accuracy of measurement analysis performed in the edge plasma.}

\proceeding{3$^{\text{rd}}$ European Conference on Plasma Diagnostics\\
  6-9 May 2019\\
  Lisbon, Portugal}
\notoc

\begin{document}
\maketitle
\flushbottom

\section{Introduction}

\begin{figure}[b]
  \centering
  \begin{tabular}{ p{0.6\textwidth} p{0.35\textwidth}}
    \includegraphics[width=\linewidth]{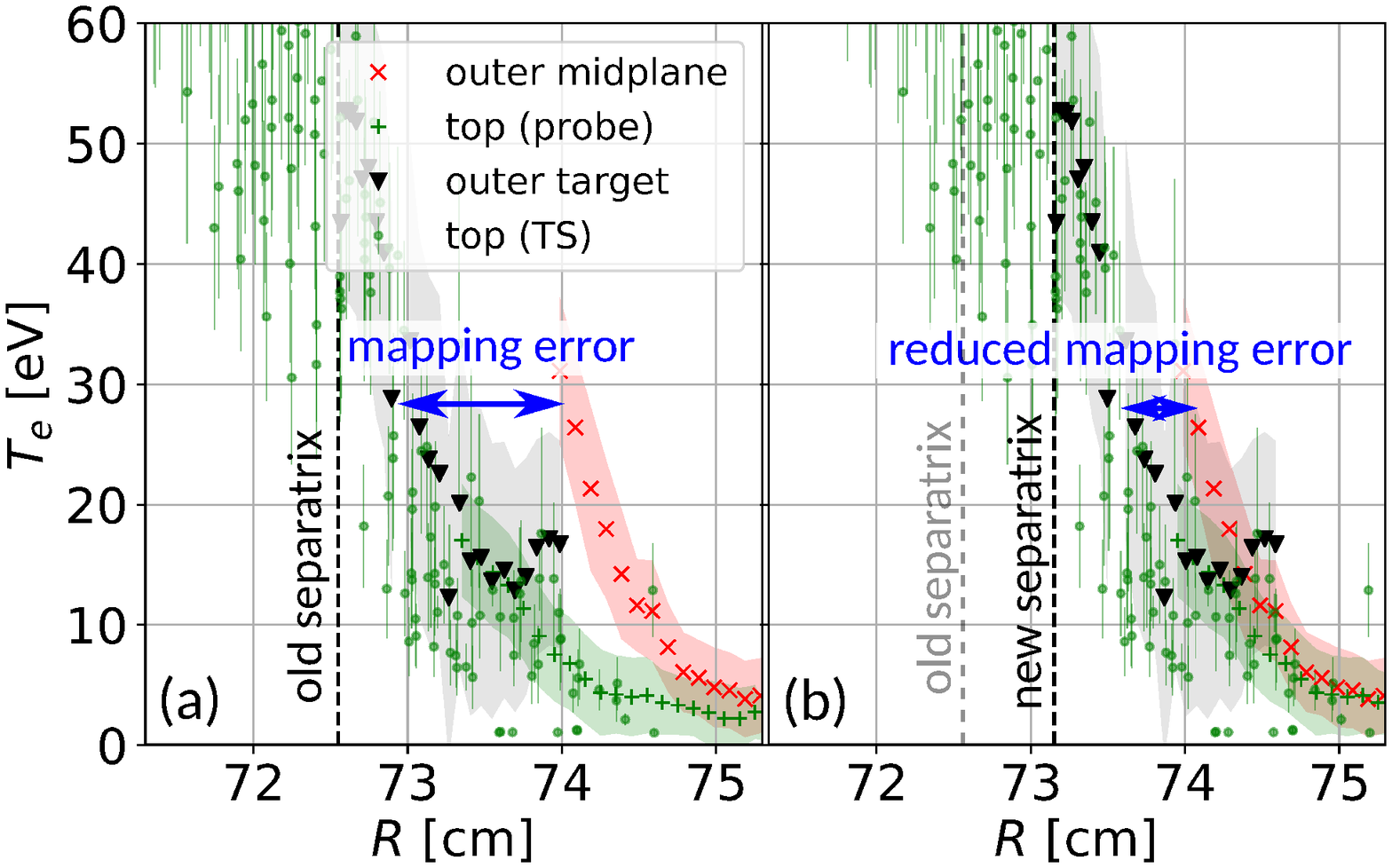} &
    \includegraphics[width=0.75\linewidth]{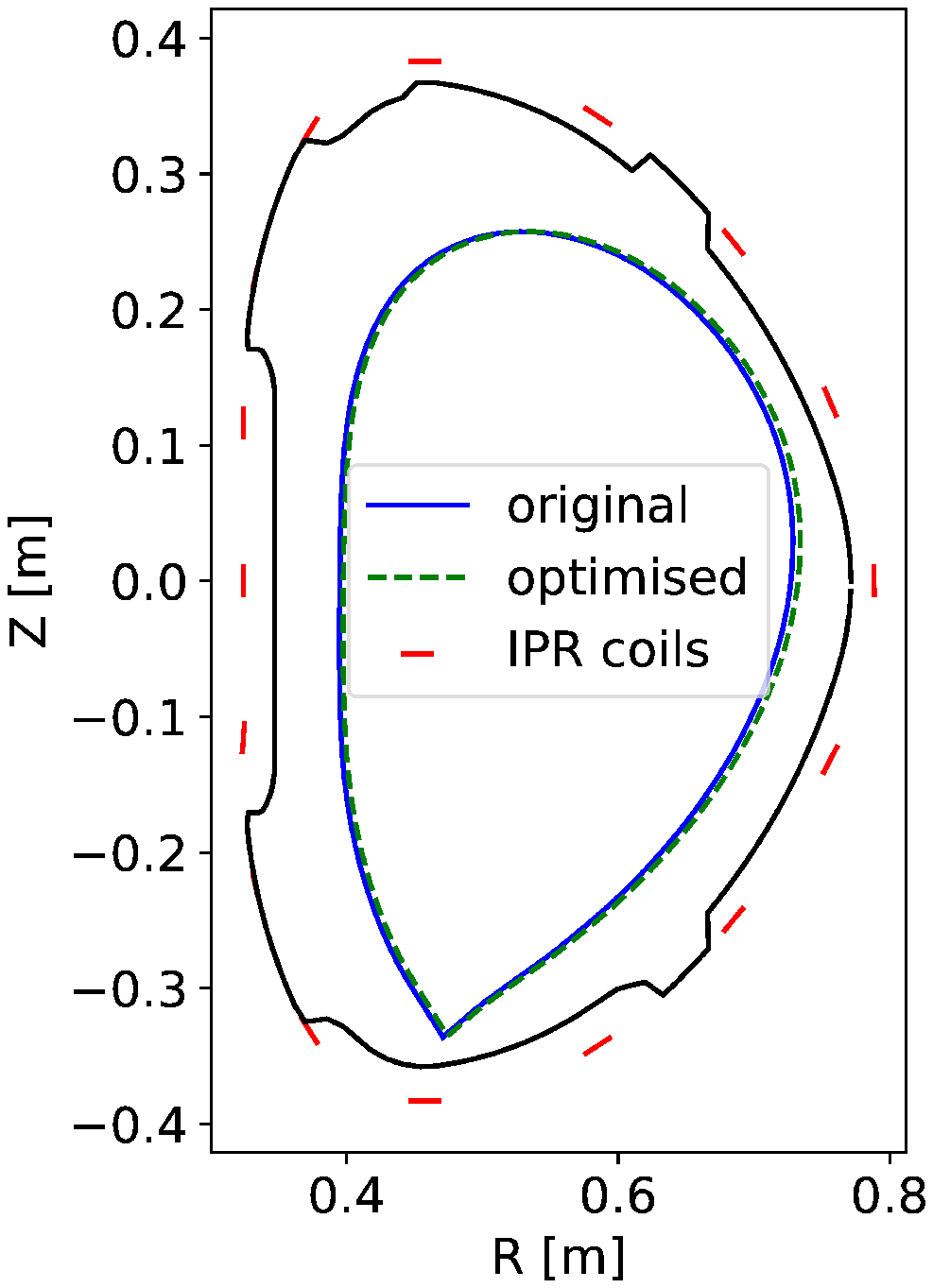}
    \\[\dimexpr-\normalbaselineskip+\abovecaptionskip]
    \caption{$T_e$ profiles measured at the OMP (reciprocating probe), the plasma top (reciprocating probe and Thomson scattering \cite{weinzettl2017}) and the outer divertor target (probe array \cite{adamek2017}), discharge \#15182. (a) Mapped to the OMP using the original EFIT reconstruction, (b) mapped to the OMP using the new reconstruction (Sec. \ref{sec:efit}).} \label{fig:mapping}&
    \caption{Location of IPR coils providing input data to EFIT, and reconstructions of discharge \#15182 separatrix using original and optimised coil positions.} \label{fig:ipr}
  \end{tabular}
\end{figure}

Magnetic equilibrium reconstruction is a vital component in interpreting experimental data collected in the tokamak edge region. However, many tokamak studies have reported problems in interpreting edge data caused by uncertanties in the equilibrium reconstruction (for example \cite{kallenbach2004, kocan2015, labombard2011, tsui2017, 
b1}).
The COMPASS tokamak is no exception. One of the known issues of its EFIT++ code reconstructions is the uncertainty of separatrix position at the outer midplane (OMP), whose effect on mapping profiles to the OMP is illustrated in figure \ref{fig:mapping}a.
It has been attempted to correct the reconstruction errors by calibrating the edge measurement position against the velocity shear layer (VSL) position instead of the EFIT separatrix \cite{seidl2017, brotankova2009}. However, efforts to address the issue on the reconstruction level have until recently been lacking.

In a companion work \cite{kovanda2019}, Kovanda et al put forth that reconstruction uncertainties may be affected by inaccurate records of the measuring magnetic coil positions in EFIT input, and they subsequently provide corrected coil positions (detailed in section \ref{sec:efit}). In this article, we systematically benchmark the resulting "corrected" reconstructions by comparing the reconstructed OMP separatrix position to the VSL position detected by electrostatic probes in L-mode. Our goal is to demonstrate systematic differences between the separatrix and the VSL position in the old reconstructions, to show that their disparity depends mainly on plasma geometry, and to corroborate that correcting for the measuring coil geometry substantially reduces the dependency. In conclusion, we recommend that equilibrium reconstructions in all previous COMPASS discharges be retroactively recalculated using the corrected measuring coil positions, as the resulting reconstructions are likely more accurate than the ones in use today.

\section{Methods}

\subsection{EFIT++}
\label{sec:efit}

EFIT++ is a standard solver of the Grad-Shafranov equation \cite{appel2001}. In reconstructing COMPASS equilibria, local magnetic fields measured by 16 inner partial Rogowski coils (IPR coils) are provided to it as minimal constraining input.\footnote{Refer to \cite{kovanda2019} for discussion on additional constraining input and reconstruction settings in COMPASS EFIT.} The IPR coils are small measuring coils distributed poloidally around the chamber (figure \ref{fig:ipr}). It was recently found that their positions recorded within the EFIT input are inaccurate. \cite{kovanda2019} Synthetic coil signals calculated by the Biot-Savart law from poloidal field coil currents in the static phase of a vacuum discharge were compared to the measured coil signals, which betrayed disagreements up to $\sim$10 \%. To infer the coil positions more accurately, their $R$ and $Z$ coordinates and the poloidal angle $\theta$ were varied so as to achieve a fit between the measured and the calculated coil signal in each individual coil. The match was found to be especially sensitive to coil angles, which were on average corrected by several degrees (not visible in figure \ref{fig:ipr}). Providing EFIT with the corrected coil positions can alleviate the mapping problems (figure \ref{fig:mapping}b).

\subsection{Velocity shear layer}

The VSL is a region in the edge plasma where the poloidal plasma velocity $v_p$ varies rapidly in the radial direction. The VSL has been shown to affect the magnitude of cross-field transport by regulating the level of plasma turbulence \cite{saha2003} and contributing to the L-H transition \cite{tynan2016}. The origin of a steady-state VSL may be, in the first approximation, connected to the transition between closed and open magnetic field lines. As argued in \cite{jirakova2018}, the interplay between the radial force balance (closed field lines) and the sheath potential drop (open field lines) causes the plasma potential $\Phi$ to peak near the separatrix, which results in a profile in the radial electric field $E_r = -d\Phi/dr$ and in the poloidal velocity $v_p = E_r\times B_t$ -- that is, a VSL. This argumentation is rather crude, but despite that a $\Phi$ peak, or the corresponding $E_r=0$, has been observed in experiment \cite{tynan2016, rognlien1999}, gyrofluid turbulence simulations \cite{manz2015}, fluid simulations \cite{rozhansky2001} and continuum kinetic simulations \cite{dorf2018} alike.\footnote{The exact relation of the separatrix and the VSL position is currently unknown --- some studies suggest that the VSL forms 0.5-1 cm outside the separatrix \cite{tsui2017, rognlien1999, manz2015, dorf2018, nold2010} while others place it up to 1 cm inside the separatrix \cite{rozhansky2001, xu2009}. It is likely that their relative position depends on a number of factors, including the connection length, plasma collisionality, attachement/detachment and more. However, section \ref{sec:results} shows that, in original COMPASS reconstructions, $R_{sep} - R_{VSL}$ is dominated by geometric factors rather than plasma parameters, reaching values from $-3$ to $+2$ cm as opposed the considerably smaller numbers found in literature. And since this systematic dependency is substantially suppressed by correcting the coil positions, which is a purely geometric adjustment, we can surmise that EFIT input inaccuracies impact the interplay between the reconstructed separatrix and VSL position significantly more than physical mechanisms.} In this paper, we exploit the fact that COMPASS probes routinely record a $\Phi$ peak to carry out a statistical comparison of the VSL position to the magnetically reconstructed separatrix position.

\subsection{Probes}

The OMP reciprocating probe of the COMPASS tokamak \cite{weinzettl2017} carries a ball pen probe, which is a Langmuir probe variation similar to the ion-sensitive probe both in design and measurement \cite{adamek2004}. Its floating potential is close to the plasma potential, 
$V_{BPP} = \Phi - \left( 0.6 \pm 0.3\right)T_e$, and for the purposes of this article we assume them equal.
A single ball-pen probe can detect the VSL centre ($\Phi$ peak) with a spatial uncertainty $\pm 2$ mm accounting for the smoothing and the neglected $T_e$ contribution.

\section{Results}
\label{sec:results}

In this section we present a statistical comparison of the EFIT separatrix radial position $R_{sep}$ to the VSL position $R_{VSL}$, carried out over a database of 398 COMPASS discharges (53 circular, 19 elongated and 325 D-shaped plasmas).
We investigate the difference $\Delta R = R_{sep} - R_{VSL}$.

\begin{figure}[t]
\includegraphics[width=\textwidth]{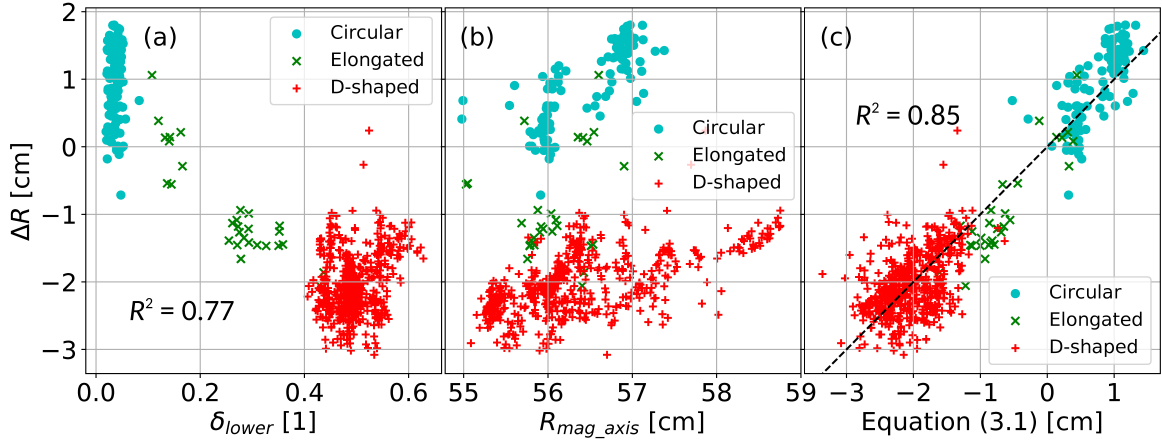}
\caption{Dependency of $\Delta R= R_{sep} - R_{VSL}$ on (a) lower triangularity and (b) magnetic axis radial position, (c) linear regression using equation \eqref{eq:delta_original}. Original EFIT reconstructions.}
\label{fig:ECPD_geometry}
\end{figure}

\begin{table}[b]
\caption{Coefficients of $\Delta R$ linear fit using 5 largest principle components of the independent variable phase space: safety factor, magnetic axis radial and vertical position, elongation, upper and lower triangularity, plasma current, toroidal magnetic field, normalised beta, and the line-averaged plasma density.}
\label{tab:PCA}
\centering
\begin{tabular}{|c|cccccccccc|}
\hline
EFIT & $q_{95}$  & \hspace{-4mm} $R_{mag\_axis}$ & \hspace{-3mm} $Z_{mag\_axis}$ & \hspace{-2mm} $\varepsilon$ & \hspace{-3mm} $\delta_{upper}$ & $\delta_{lower}$ & \hspace{-3mm} $I_p$ & \hspace{-3mm} $B_t$ & $\beta_N$ & $\overline{n}_e$\\
\hline
original & $-0.1$ & \hspace{-4mm}\textbf{0.5} & \hspace{-3mm}$\bm{-0.3}$ & \hspace{-2mm}0.09 & \textbf{0.3} & \hspace{-3mm}$\bm{-1.3}$ & \hspace{-3mm} $-0.03$ &  \hspace{-3mm}0.03 & \textbf{0.4} & 0.01 \\
new & $-0.02$ & \hspace{-4mm}\textbf{0.2} & \hspace{-3mm} 0.06 &\hspace{-2mm} 0.1 & \hspace{-3mm} $\bm{-0.4}$ & $-0.03$ &\hspace{-3mm} 0.08 & \hspace{-3mm} $-0.07$ & \textbf{0.2} & $-0.007$ \\
\hline
\end{tabular}
\end{table}

\begin{figure}[t]
\includegraphics[width=\textwidth]{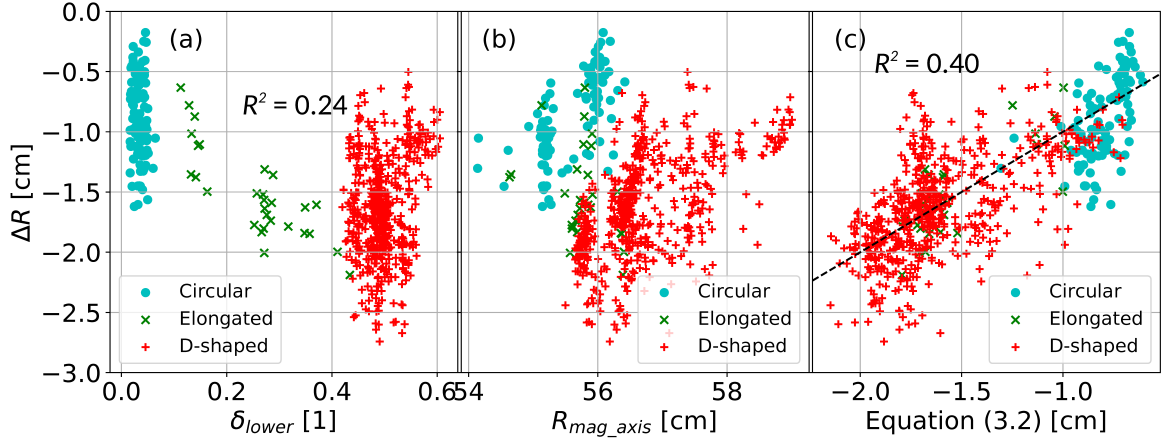}
\caption{Dependency of $\Delta R= R_{sep} - R_{VSL}$ on (a) lower triangularity and (b) magnetic axis radial position, (c) linear regression using equation \eqref{eq:delta_optimised}. New EFIT reconstructions.}
\label{fig:ECPD_geometry_optimised}
\end{figure}

Figure \ref{fig:ECPD_geometry}a shows that $\Delta R$ varies considerably across the COMPASS database, from -3 cm to +2 cm, and that this variation consists of a random component and a  systematic component.
To find which variables affect the systematic component, we evaluated the dependence of $\Delta R$ on the variables listed in Table \ref{tab:PCA} using the Principle component analysis (PCA). We found the 5 largest principle components of the phase space, responsible for 91\% of its variance, and with them acting as the independent variables we performed a linear regression of $\Delta R$. The regression matched closely with the data, $R^2=0.86$. Subsequently, we transformed the principle components back into the variables of Table \ref{tab:PCA}, obtaining the coefficients listed in Table \ref{tab:PCA}. In the original EFIT reconstructions, $\Delta R$ is observed to depend most strongly on the plasma lower triangularity $\delta_{lower}$ and the magnetic axis radial position $R_{mag\_axis}$, which both relate to the plasma geometry. In figures \ref{fig:ECPD_geometry}a and \ref{fig:ECPD_geometry}b, one may observe both the dependencies.  Finally, figure \ref{fig:ECPD_geometry}c shows the aforementioned linear regression with "reduced" variables - only the emboldened coefficients in Table \ref{tab:PCA} were considered, that is,
\begin{equation}\label{eq:delta_original}
\Delta R = -1.6 - 1.3\delta_{lower} + 0.5R_{mag\_axis} + 0.5\beta_N - 0.3Z_{mag\_axis} + 0.3\delta_{upper}
\end{equation}
with an almost unchanged $R^2=0.85$.
In figure \ref{fig:ECPD_geometry_optimised}, the same plots are presented for the new EFIT reconstructions. We see that purely geometrical adjustments to the EFIT input have a major impact on the reconstructed separatrix position. As observed in Table \ref{tab:PCA}, some dependency on triangularity and the magnetic axis position remains, but it is much less pronounced compared to the random error.
The "reduced" linear regression of $\Delta R$ for the optimised EFIT is
\begin{equation}\label{eq:delta_optimised}
\Delta R = -1.5  + 0.2 \beta_N + 0.2 R_{mag\_axis} - 0.4\delta_{upper},
\end{equation}
with $R^2=0.4$, which shows a significant suppression of the systematic component of $\Delta R$.

\section{Discussion and conclusions}
\label{sec:discussion}

We have compared the outer-midplane position of the magnetically reconstructed separatrix $R_{sep}$ to the velocity shear layer (VSL) position $R_{VSL}$ and drawn two conclusions: (i) current EFIT reconstructions contain a systematic error dependent on plasma geometry, and (ii) this error can be mitigated by correcting magnetic coil positions recorded in the EFIT input. It should be mentioned that although in previous works the VSL has been consistently associated with the separatrix, it is true that they may not coincide. In original COMPASS reconstructions, nevertheless, $R_{sep}-R_{VSL}$ is dominated by geometric factors to the point where other physical dependencies are relatively inconsequential. We thus recommend using the corrected coil positions as a solid step toward more reliable and accurate equilibrium reconstructions in COMPASS. Using this experience, similar problems can be avoided in the future COMPASS-Upgrade tokamak equilibrium reconstructions.

\acknowledgments

This work was carried out within the framework of the EUROfusion Consortium and has received funding from the Euratom research and training programme 2014-2018 and 2019-2020 under grant agreement No 633053. The views and opinions expressed herein do not necessarily reflect those of the European Commission. This work was supported by the Grant Agency of the Czech Technical University in Prague, grant No. SGS19/180/OHK4/3T/14, project No. CZ.02.1.01/0.0/0.0/16\_019/0000768, Czech Science Foundation grant No. GA19-15229S, and co-founded by MEYS projects number 8D15001 and LM2015045.

\end{document}